\documentclass[12pt]{article}
\usepackage[utf8]{inputenc}
\usepackage[T1]{fontenc}
\usepackage{titling}
\usepackage{float}
\usepackage{comment}
\usepackage{tikz}
\usepackage[none]{hyphenat}
\usepackage{url}
\usepackage[colorlinks,citecolor=blue,urlcolor=black]{hyperref}
\usepackage{multicol}

\usepackage[sorting=none, style=nature]{biblatex} 
\bibliography{references}

\usepackage{anysize}
\usepackage{authblk}

\marginsize{2cm}{2cm}{2cm}{2cm}

\title{New Horizons: Pioneering Pharmaceutical R\&D with Generative AI from lab to the clinic - an industry perspective}
\author[1, 4]{Guy Doron}
\author[2, 4]{Sam Genway}
\author[2,3]{Mark Roberts}
\author[1]{Sai Jasti}
\affil[1]{Data Sciences \& AI, R\&D, Pharmaceuticals, Bayer AG, Berlin, Germany}
\affil[2]{Hybrid Intelligence, Capgemini Engineering, Stevenage, United Kingdom}
\affil[3]{Generative AI Lab, Capgemini, Paris, France}
\affil[4]{Equal contribution}

\begin{document}

\maketitle

\begin{titlepage}

\end{titlepage}

\begin{abstract}

The rapid advance of generative AI is reshaping the strategic vision for R\&D across industries. The unique challenges of pharmaceutical R\&D will see applications of generative AI deliver value along the entire value chain from early discovery to regulatory approval.
\\

This perspective reviews the challenges in pharmaceutical R\&D and takes a
\linebreak
three-horizon approach to explore the generative AI applications already delivering impact, the disruptive opportunities which are just around the corner, and the longer-term
\allowbreak
transformation which will shape the future of the industry. Selected applications are reviewed for their potential to drive increase productivity, accelerate timelines, improve the quality of research, data and decision making, and support a sustainable future for the industry. Recommendations are given for Pharma R\&D leaders developing a generative AI strategy today which will lay the groundwork for getting real value from the technology and safeguarding future growth.
\\

Generative AI is today providing new, efficient routes to accessing and combining organisational data to drive productivity. The technology is already advancing the design of novel small-molecule drugs and driving new machine learning applications through the generation of synthetic data. Next, this impact will reach clinical development, enhancing the patient experience, driving operational efficiency, and unlocking digital innovation to better tackle the future burden of disease. Looking to the furthest horizon, converging technologies will transform pharmaceutical R\&D. Rapid acquisition of rich multi-omics data, which capture the “language of life”, in combination with next-generation AI technologies will allow organisations to close the loop around phases of the pipeline through rapid, automated generation and testing of hypotheses from bench to bedside. This provides a vision for the future of R\&D with sustainability at the core, with reduced timescales and reduced dependency on resources, while offering new hope to patients to treat the untreatable and ultimately cure diseases.

\end{abstract}


\newpage

\section{Introduction}
\label{Section:Introduction}

\subsection{Meeting the challenges of pharmaceutical R\&D}
It is only through years of pharmaceutical R\&D that new medicines are discovered, tested, and approved to reach patients \cite{REF1_wouters2020estimated, REF2_singh2023drug}. The journey from conceptualisation to a novel therapeutic is driven by scientific research and innovation, an ethical commitment to advancing global healthcare in an equitable and sustainable way, and a determination to respond to unmet healthcare needs. Data-driven decision making is central to tackling many of the industry challenges, and AI is already seeing an impact. Throughout the industry, novel therapeutics face a series of challenges across the value chain along the route to the patient. These headwinds arise from the long lifecycles for R\&D today, the substantial end-to-end costs, and the challenging quest for novel routes to treating the untreatable.
\\

As a relative newcomer to the scene, generative AI technology (BOX 1) is already disrupting the traditional approaches to many tasks within the R\&D lifecycle. Beyond the current horizon lies a broad portfolio of opportunities to improve the speed of R\&D, enhance productivity and control cost, and enable high-quality research – all while supporting long-term strategies for the sustainability of the industry, safeguarding a vision of healthcare for all.
\\

This pharma industry perspective presents an ambition for the application of generative AI technologies across pharmaceutical R\&D, which centres on their impact on enhanced \textit{speed, productivity, quality} and \textit{sustainability}. Before giving a three-horizon view of the specific opportunities the technology offers, a review of the landscape of challenges is summarised, and the methodology for exploring new applications is given. 

\subsection{The long haul for R\&D}
The journey of a new medicine involves many more setbacks than successes. In today's landscape, bringing a new medicine to the market requires an investment on the order of \$1 billion and typically takes well over a decade \cite{REF1_wouters2020estimated, REF2_singh2023drug}. From the inception of a potential drug concept to its eventual distribution among patients, this timeline involves distinct phases which can be broadly grouped into: (i) identifying a novel biological target, (ii) finding a novel drug candidate, (iii) preclinical assessments, (iv) clinical trials and (v) regulatory approval.
\\

Taking into account the cost of failures, typically a little under half of the investment and time is spent during the discovery stages with the rest during preclinical and clinical development \cite{REF3_paul2010improve}. As well as aspiring to reach patients with new treatments quickly, the lifetime of patents – typically 20 years – means increased development times can reduce the ultimate profitability of a novel therapeutic. The daunting reality is that a significant number of promising drug candidates falter at various stages of development; many stumble before leaving the preclinical phase, while less than 10\% get through clinical trials \cite{REF4_REF7_sun202290}.
\\

This high attrition rate is more than a statistical footnote; it underscores the need for an agile and robust target validation, predictive preclinical models, and the need to distinguish viable candidates from potential dead-ends in a fail-fast-and-learn culture. Applications of generative AI which focus on delivering speed and productivity to stages in the R\&D pipeline will be key to shortening timelines and reducing costs. Similarly, use cases which improve the quality of data, assets and decision-making will play a pivotal role in reducing the failure rates across the phases of R\&D. One of the topics this perspective will explore is where these opportunities are unfolding today, from facilitating rapid access to organisational knowledge to generative design molecular design.

\subsection{Treating the untreatable}
The landscape of treatments is experiencing a shift as scientific discoveries have unveiled a host of new treatment modalities and the role of AI in drug discovery has grown. Year-on-year, the proportion of biologics in new approvals is increasing \cite{REF5_senior2023fresh}, with cell therapies, gene therapies, nucleic acid therapeutics and other new modalities all gaining a share of the pipeline \cite{REF6_blanco2020new}. Beyond offering the potential for new selective and potent drugs, new modalities and AI-first drug discovery are expanding the druggable target space, and increasing the likelihood that new targets found will be addressable. Increasingly, the challenge of moving from scientific breakthroughs to new treatment will require collaboration not only between experts across biology, chemistry and engineering, but also those in data science and AI.
\\

Finding new druggable targets through identifying the biological processes that can be modulated by medications to achieve a therapeutic effect is core to conceptualising a new treatment. While the horizon of potential targets expands with each new discovery, discovery and development can fail if it is not possible to identify safe and efficacious therapeutic \cite{REF4_REF7_sun202290}. Indeed, for first-in-class drugs, the ultimate validation of a target may be only possible at the clinical stage, something which is responsible for a significant proportion of late-stage failures \cite{REF8_hingorani2019improving}. The process of target identification requires an understanding of disease mechanisms and underlying biological pathways. This is amenable to generative AI techniques through the digestion of scientific literature which can make the knowledge accessible, a topic which will be developed in this perspective in the context of improving scientific productivity and accessing higher \textit{quality} data.
\\

Another dimension is developing treatments for rare diseases where, necessarily, the economic implications of developing a product for small number of potential patients mean making the case for R\&D investment is difficult. Despite incentives \cite{REF9_EMAOrganDesignationOverview}, there are more than 7000 identified rare diseases for which only around 500 have approved treatments \cite{REF10_NIHPromiseOfPrecisionMedicineRareDiseases}. R\&D strategy must focus on identifying routes to developing high-performing therapeutics for smaller patient subpopulations in a drive towards precision medicine. This perspective will discuss how
\allowbreak
generative AI will play a role through being able to learn complex representations from 
\allowbreak
high-dimensional patient data which facilitate the use of synthetic data in modelling disease subpopulations, which could eventually lead to cures for rare diseases.

\subsection{Sustainable health for all}

In a world increasingly interwoven by technology, healthcare disparities across the globe are still marked. Communities around the globe remain bereft of essential medications, emphasising the importance of developing global approaches to affordability and accessibility. It is important that diverse global population is represented in the development of new medicines and that revolutionary therapies find their way to all corners of the world. The focus above on generative AI to meet the needs of smaller subpopulations and reduce costs while enabling broader representation of the global population will have significant impact on future medicines.
\\

As demographics shift, the aging population is reshaping our perception of growing old. By 2050, the number of people aged over 60 worldwide is anticipated to double, reaching a staggering two billion \cite{REF11_WHOAgeingAndHealth}. 
The resulting increase in age-related ailments such as dementias, cardiovascular diseases, and cancers will drive and increasing demand for new interventions that enhance both longevity and human quality of life. The imperative is to address not only the treatment of these conditions but also the innovative ways in which healthcare can support the aging population. This perspective will discuss how generative AI can play a role in new approaches and offer hope in enabling individuals to lead not only longer, but healthier, happier lives.
\\

More broadly, the evolution of sustainable pharmaceutical R\&D is now at the centre of corporate strategy, motivated not just by company responsibility but also futureproofing their businesses for longer-term growth. Sustainability is multifaceted and with impact across the value chain and a \textit{sustainability} dimension will be considered for the use cases discussed in this perspective. This dimension is particularly important: while new generative AI applications will have a significant impact across the value chain, they will come with an environmental cost due to their energy consumption. Despite algorithmic and hardware advances which are driving efficiencies \cite{REF12_desislavov2023trends}, the proliferation of new generative AI applications could have significant global impact. It will therefore be important to explore the multitude of generative AI applications which drive sustainability through operational efficiencies and increased productivity, and continually explore approaches which minimise the environmental impact \cite{REF13_HBRHowToMakeGenerativeAIGreener}.

\subsection{A new era of innovation from converging technologies}

Within the shifting landscape of challenges, a remarkable convergence of cutting-edge
\linebreak
technologies is taking centre stage. The ensemble comprises advances in gene editing, omics, multiplex screening, bioengineering, diagnostics, wearables, and the transformative potential of generative AI. This collective force evolving the landscape of drug discovery, charting a future for personalized medicine, and shaping the future of clinical trials.
\\

\begin{figure}[H]
    \centering
    \includegraphics[scale=0.6]{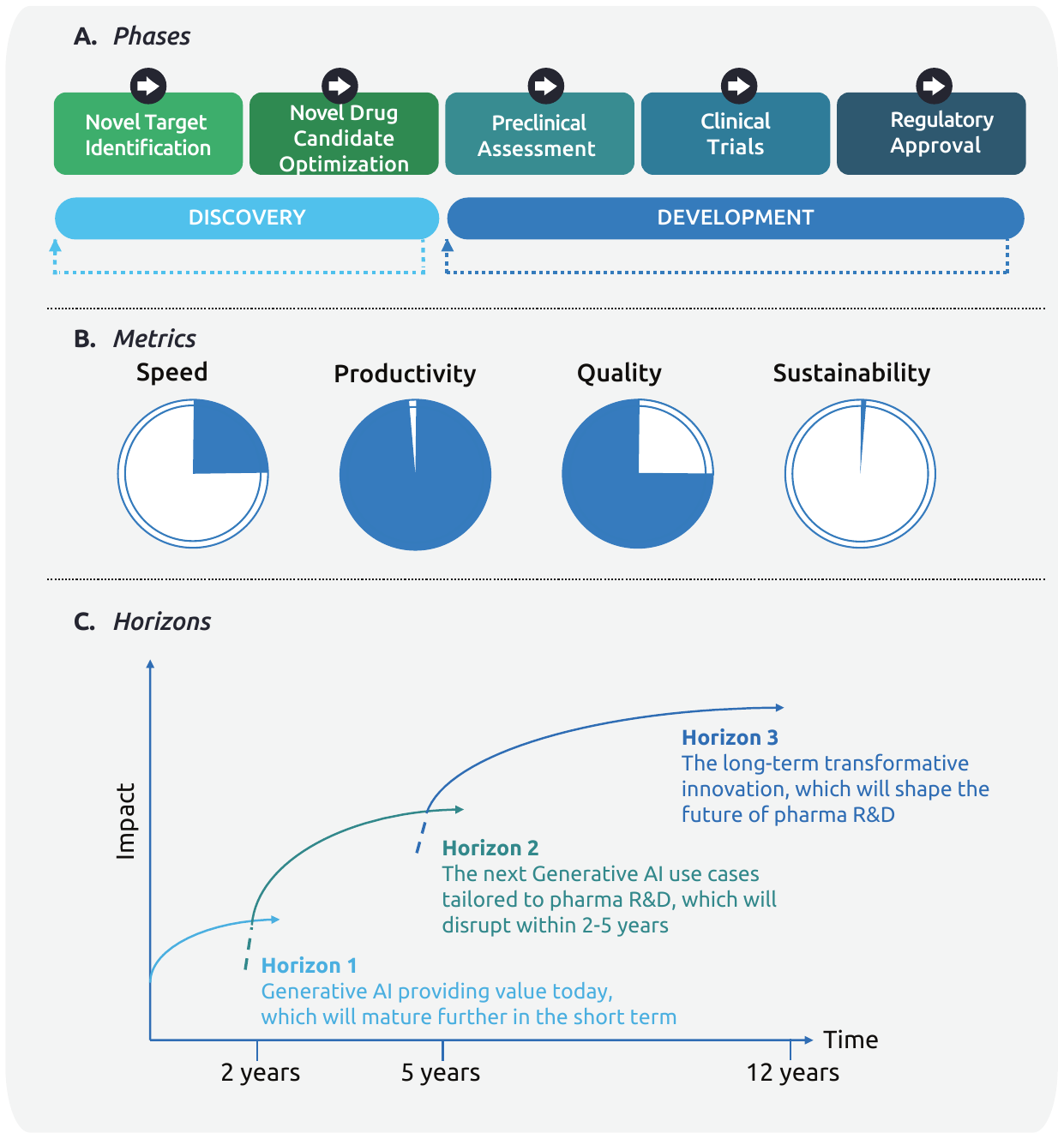}
    \caption{A model for future development and integration of generative AI into the pharmaceutical R\&D organisation to provide a vision for the future of the technological impact across the value chain. This model illustrates the three horizons of innovation in pharma R\&D, along with their associated phases, metrics, and impact.}
    \label{fig:Horizonsfig1}
\end{figure}

At the heart of this convergence, generative AI emerges as a transformative force, poised to impact not only our present and immediate future, but also the distant tomorrow. It brings renewed hope for patients, offering innovative treatments and potentially paving the way towards curing diseases. As well as looking at the opportunities for generative AI through the lenses of \textit{speed, productivity, quality} and \textit{sustainability}, a three-horizon view will be developed to look at opportunities in each of the stages from the lab to the clinic (Figure \ref{fig:Horizonsfig1}). The next section will focus on Horizon 1 - the use cases which can deliver value today. Following this, Horizon 2 will look into the next opportunities visible in the near horizon, reaching maturity within 2--5 years. A vision for the longer-term third horizon of 5--12 years will be described thereafter. The concluding section will then outline some suggested next steps.


\centerline{
\fbox{
\begin{minipage}{\linewidth}
\centerline{
    \large Box 1: What is generative AI?
    \label{BOX1}
}

\vspace{2em}
Generative Artificial Intelligence (AI) is focused on the development of AI models capable of producing data autonomously, sometimes mimicking human expertise. This includes generating text, images and even molecular structures. Unlike traditional AI approaches, which primarily focus on classification and prediction, generative AI can support tasks ranging from creative design to language translation, summarisation, question-answering and even writing source code in various programming languages. To understand the scale of this progress, an artistic example is helpful: while traditional AI can identify if a piece of music was written by Mozart, generative AI can write a new composition that sounds just like Mozart. Although generative AI is far from new, it has recently reached a tipping point in public awareness and, following impressive performance on a range of tasks, it now features in many organisations’ future strategies.
\\

At the core of all generative AI is the use of deep learning algorithms to learn complex patterns in what are often huge volumes of complex data. This enables it to generate new data of the same form, whether images, text or biological sequence data. Crucially, generation can be conditioned on inputs to the model. For example, this may be to generate pictures of a particular scene specified by a description, rather than generating random images. In the case of text, this may mean generating words which follow from a set of words given as input, for example to complete a sentence or respond to a query. This is known as \textit{prompting}.
\\

The concepts around how generative AI learns patterns in complex data are common across a range of generative AI model architectures. In large, complex datasets, there is typically a great deal of structure: for example, there are many ways to put words together to make a coherent sentence, but almost all random collections of words will not make a real sentence because they lack grammatical structure. Similarly, there are vast numbers of convincing images an AI model could generate, but choosing image pixels at random will almost never generate a real-looking image, because real images contain objects which have structure.
\\

Of the many model architectures for generative AI, large-scale transformer-based models such as generative pretrained transformers (GPT) have gained particular notoriety for their performance in generating coherent and context-rich text. Such large language models are frequently known by the abbreviation LLMs.

\end{minipage}
}
}

\section{Horizon 1 opportunities for generative AI - The value for pharmaceutical R\&D today}
\label{Horizon1Opportunities}

Several opportunities for generative AI are already delivering value to R\&D or will reach maturity in the very near future. The impact on \textit{speed, productivity, quality} and \textit{sustainability} spans across use cases, with relevance to many tasks along the value chain. In the near term, generative AI will contribute to accelerating the time to reach discovery and development milestones, increasing the quality of targets, candidate molecules and decisions made, and increasing productivity of the R\&D workforce. This section reviews these opportunities and qualifies their impact (Figure \ref{fig:Horizon1fig2}).
\\

\subsection{Accessing organisational knowledge: the virtual assistant}

The ability of conversational agents built on large language models (LLMs), such as ChatGPT, 
to answer questions from a range of users has prompted myriad applications which promise to drive productivity in staff across organisations. The LLM can serve as a gateway to organizational knowledge, enabling effortless access to relevant information stored within
\allowbreak
internal databases and documents. This rapid accessibility reduces the time it takes to search for information and will ultimately offer the ability to make more informed decisions. If it is possible to get answers within seconds which would otherwise take days of research, accessing knowledge and will become a routine part of daily decision making (Figure \ref{fig:Horizon1fig2}, `Virtual Assistant').
\\

Because LLMs are prone to responding to human prompts with confident answers even when their answers are wildly incorrect - a phenomenon known as hallucination – it is necessary to integrate LLMs with datasets which are relied upon to source answers 
\cite{REF14_MicrosoftInsightsFromDataAzureOpenAI}. 
Approaches such as retrieval augmented generation (RAG) offer a dynamic approach to data retrieval and analysis.
By connecting with internal databases and repositories, LLMs can extract and analyse data, offering valuable insights and streamlining data-driven processes. Crucially, where LLMs leverage existing data sources, it is possible to verify the answers given by the LLM virtual assistant by tracing the response back to the source data.
\\

Additionally, LLMs are offering support in task automation and integration. They perform a wide range of functions, which can include automating routine tasks like summarising or translating textual information, even at the level of email correspondence. Another area where LLMs offer near-term value is in the automation of report and document creation, which has particular relevance in the clinical phases of drug development \cite{REF15_mesko2023imperative}. While human review will be essential, the increase in productivity from automating generation could be significant. Beyond these established use cases, an LLM-based virtual assistant can connect with additional services, which may be specialised AI models hosted within an organisation, or external APIs. This flexibility is beginning to allow them to conduct complex tasks, such as supporting scientists in the optimisation and synthesis of drug molecules, and example where utility has already been demonstrated \cite{REF16_bran2023chemcrow}. 
\\

Despite rapid progress in this area, it is essential to acknowledge the challenges and risks associated with implementing LLMs in this context. Ensuring data privacy and security, mitigating biases in AI-generated responses, and managing user expectations are crucial
\allowbreak
considerations for certain use cases. Beyond this is the challenge of sustainability: LLMs have a notable per-query energy cost \cite{REF17_samsi2023words} and should be used judiciously to drive efficiencies which balance this environmental cost. Organizations must strike a balance between harnessing the benefits of LLMs and addressing these challenges to fully leverage their potential for organizational productivity. Part of this will come through education and awareness about the strengths and weaknesses of the technology.
\\

\subsection{Harmonising datasets to find targets and patient subpopulations}

Generative AI can aid in the exploration of novel drug targets by efficiently providing insights from vast scientific literature -  see (Figure \ref{fig:Horizon1fig2}, `Targets \& Patient Subpopulation Identification'). It can consume papers to uncover genes or proteins with emerging disease links, revealing potential connections that could be obscured by the sheer volume of available information. These insights can guide researchers towards promising candidates, which are then validated through experiments, underscoring the collaborative synergy between AI-driven literature
\allowbreak
analysis and human expertise in expanding the landscape of drug target identification.
\\

While training LLMs on biomedical data is a natural route forward, impressive performance has been derived from the combination of large language models with web APIs, facilitating access to public data \cite{REF18_REF19_jin2023genegpt} for a range of applications including drug target identification. Other routes involve cutting-edge relation-mining pipelines to create knowledge graphs (KGs) at scale with reasonable accuracy  \cite{REF18_REF19_jin2023genegpt}, opening the door to new discoveries which would be inaccessible through manual literature search. Indeed, the tandem of LLMs and KGs presents an important current direction, particularly in the realm of target discovery. KGs play a vital role in this context by efficiently integrating heterogeneous information and complex relations within biological systems. Multiple approaches are emerging to help create systems which can respond and reason better though leveraging the rich factual information in KGs with the expressibility of LLMs to enable systems with better reasoning capability \cite{REF20_pan2023unifying}.
\\

Identifying the patient subpopulations which will benefit from a novel therapeutic is another key challenge which can be addressed through digesting and building representations of complex data. The move to precision medicine is enabled though learning the structure of disease populations, for example through digesting real-world data about patients. Generative AI techniques can allow latent representations of the data to be created, which enable the structure of subpopulations to be elucidated \cite{REF21_pratella2021survey}.
\\

\subsection{Designing drug molecules – small at first}

LLMs have found domain-specific utility by extending their capabilities beyond natural language to the intricate "language" of chemistry and biology. An illustrative instance is the utilization of Simplified Molecular-Input Line-Entry System (SMILES) strings, employing textual tokens to encode chemical structures. The grammatical constructs within SMILES strings offer an expressive means of representing organic molecules at the atomic and bond level. These strings encode the information needed to construct 2D molecular graphs. 
\\

Generative models can be trained on extensive repositories of SMILES strings, enabling them to generate novel and meaningful SMILES representations for molecular structures that have never been synthesized before (Figure \ref{fig:Horizon1fig2}, `Drug Molecules Design'). Techniques like fine-tuning facilitate LLMs in focusing on specific chemical spaces relevant for specific targets. Molecular LLMs have spurred various applications, including leveraging the LLM representation for predictive modelling, chemical reaction prediction, and refining molecular structures to propose new molecules containing relevant chemical fragments or features. Likewise, amino acid sequences serve as a foundational training component for LLMs
\allowbreak
focusing on proteins, with equivalent approaches learning from nucleotide sequences. Such foundation models allow a range of protein properties to be predicted \cite{REF22_chandra2023transformer}. While generative AI for de novo protein design is still in its infancy, there is value in understanding target structures from LLMs, which have excelled in generating representations which predict protein folding successfully, yielding remarkable accuracy 3D structure prediction from amino acid sequence data \cite{REF23_rao2020transformer}. Such models have further provided a foundation for tasks in other phases of R\&D, such as detecting genetic missense variants which alter sequence of residues with pathogenic effect \cite{REF24_cheng2023accurate}.
\\

The idea of using language models for generative chemistry goes back over six years before the advent of transformer architectures, leveraging recurrent neural networks \cite{REF25_segler2018generating} and attention mechanisms \cite{REF26_pogany2018novo}. Since then, significant research has been undertaken, with LLMs such as MegaMolBART which is trained on over a billion molecules \cite{REF27_MegaMolBARTRepository}, and cMolGPT which uses target-specific embeddings to generate molecules active against different targets \cite{REF28_wang2023cmolgpt}. The ability to generate large and diverse sets of novel molecules is not, on its own, enough to have an impact.  It is essential that generation can be focused towards the active molecules with the optimal properties.  As there are limits to the number of compounds which can be synthesised and assayed, this means selecting subsets to progress in the molecular design process. It is still a challenge to narrow down the set of generated molecules.  Generative chemistry is therefore most useful as part of a platform of tools which, combined with human expertise,  help to speed up drug design and identify better candidates \cite{REF29_green2020bradshaw, REF30_ivanenkov2023chemistry42}.  Similar efforts in generative AI for de novo protein design are beginning to take hold, such as antibody design \cite{REF31_xu2023ab}, and will be revisited in the next section.

\subsection{Synthetic data and drug development}

\begin{figure}[H]
    \centering
    \includegraphics[scale=0.6]{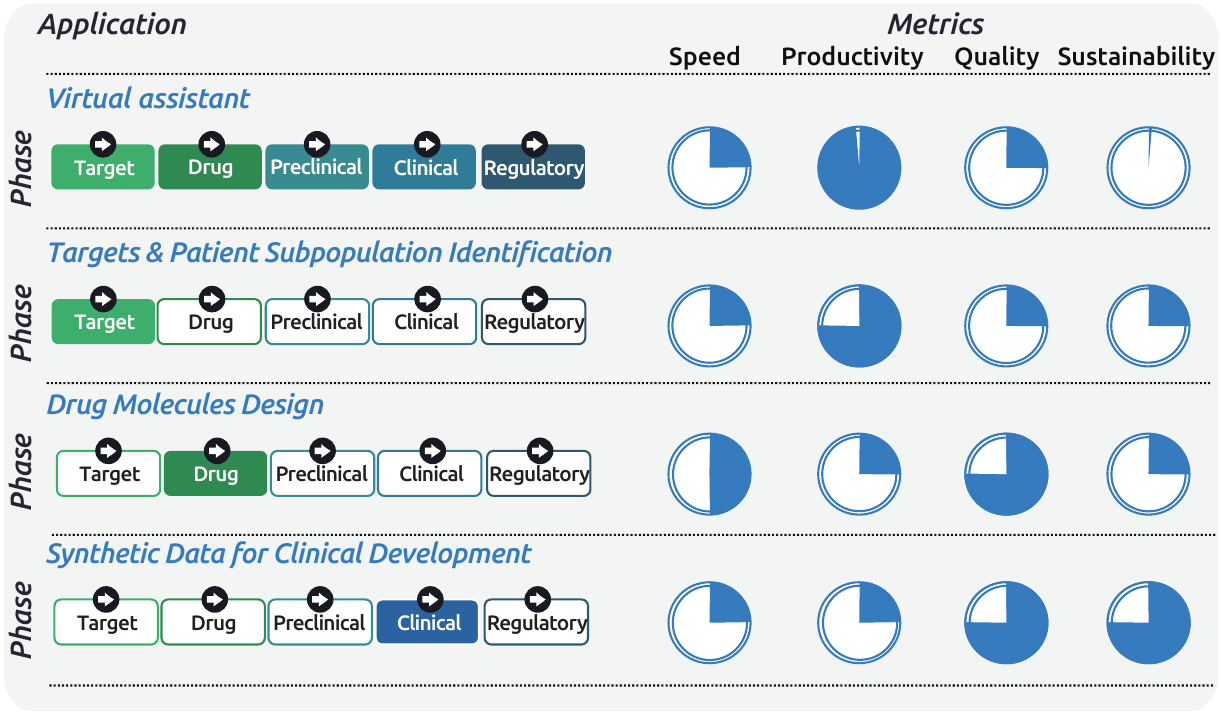}
    \caption{Horizon 1: Suggested use case development timeline, showing key stages and metrics. Speed, productivity, quality, and sustainability are key metrics for suggested use case development.}
    \label{fig:Horizon1fig2}
\end{figure}

Generative AI has long been well-suited to enriching training datasets, such as those used to model clinical trials and clinical outcomes, by producing synthetic data that mirrors the statistical attributes of actual patient data \cite{REF32_goncalves2020generation, REF33_azizi2021can} (Figure \ref{fig:Horizon1fig2}, `Synthetic Data for Clinical
Development'). The value of synthetic data in drug development has increased as the 
\linebreak
opportunities for AI have become more widespread \cite{REF34_walonoski2018synthea, REF35_krenmayr2022ganeraid}. These encompass prognostic, diagnostic, and responder models that aid in patient outcome prediction, early disease detection, and treatment personalisation. Another area is subgroup analysis, unveiling hidden patient population patterns for targeted therapies. A central benefit to using synthetic data is 
\allowbreak
overcoming data availability challenges. This includes potential issues with due to regulatory issues, especially when data is required to be shared internationally \cite{REF36_james2021synthetic}. More broadly, synthetic data can accelerate access to data to start prototyping models quickly and can provide a route to overcoming issues such as data retention, restrictions on secondary use, or collaborating with third party services providers. 
\\

Beyond enabling and accelerating access, augmenting datasets with synthetic data is also of interest. While data augmentation can be in the creation of high performing models, it is not without risk. Indeed, what it means to have statistically identical synthetic patient data requires careful definition. Particularly problematic are attempts to extrapolate outside of the domain for which data are available, and rigorous validation is essential. There are, however, situations where synthetic data holds promise, even in the high-stakes scenario of a clinical trial. One scenario is the creation of synthetic control arms leveraging real-world data from sources outside of the trial \cite{REF37_AppliedClinicalTrialsDataEquity}. Synthetic control arms are appealing from the perspective of increasing speed and the potential to support the inclusion of underrepresented groups.

\section{Horizon 2: The next opportunities for pharma R\&D}

Progress in generative AI is advancing at a remarkable pace. Today, researchers and innovators are actively exploring, developing, and rigorously evaluating the applications of tomorrow. Horizon 2 disruption comes from development of Horizon 1 opportunities towards deeper impact, such as smarter virtual assistance, alongside a new set of use cases which will emerge over the next two to five years, including generative AI playing a role in digital therapeutics (Figure \ref{fig:Horizon2fig3}). Application areas will be reviewed along the stages of the pharmaceutical R\&D pipeline, with the strategic value linked to \textit{speed, quality, productivity and sustainability}. 
\\

\subsection{The virtual assistant gets smarter}

Moving beyond the current capability of generative AI technologies to facilitate the rapid retrieval of information from across an organisation through a natural language interface, there are already signs of wider capabilities in problem solving developing \cite{REF38_davis2023testing}. Furthermore, the ability of generative AI to generate functional code in a range of programming languages is already having an impact in the real world \cite{REF39_peng2023impact}. AI which can truly reason may be a way off, but the next generation of tools will be more modular in nature and offer an interactive end-to-end capability in the exploration of scientific ideas for R\&D (Figure \ref{fig:Horizon2fig3}, `Smarter Virtual assistant'). This will be driven by the ability to leverage expert systems and specialist AI tools alongside the wealth of public and in-house data available for pharmaceutical discovery and development. Beyond being able to respond to particular queries, the researcher will be able to explore ideas end-to-end, iteratively and quickly, drawing on a broad range of tools which the virtual assistant will identify. The ability to access data, AI models and code generated will make the findings possible to reproduce and validate.
\\

Virtual assistant use cases can also break down organisational barriers. The need for effective translation of specialist jargon for different audiences exists between multiple business functions across R\&D. Indeed, there is a great deal of bespoke terminology which may be custom the scientific research in a specific phase, or to a particular therapeutic area. The impact of generative AI to break down organizational barriers which arise between domain experts could have a significant impact on productivity. The effect could be felt outside of the organisation too, enabling the market to understand the science in the research lab as well as the decision-making in the board room.
\\

\subsection{The era of generative biology}

As discussed in the previous section, generative AI applied to protein structures modelling has seen rapid development, particularly in the application of AI-driven protein folding techniques such as AlphaFold \cite{REF40_jumper2021highly}, RoseTTA fold \cite{REF41_baek2021accurate}, and ESM \cite{REF42_lin2023evolutionary}. However, the impact of these advancements on pharmaceutical discovery has not been as extensive as is often imagined from outside the industry. The focus on natural proteins analysis has meant the technology is most relevant for finding biological targets. However, predicting target protein structures with AI is not always fundamental. Indeed, target structures may already be known from crystallography, or it may be that phenotypic approaches can make progress without knowledge of a target structures. Beyond these reasons, even if generated protein structures are useful, capturing protein dynamics or changes to the structure in presence of drug molecules may be important \cite{REF43_NatureAlphaFold}.
\\

Since these initial breakthroughs, recent and rapid progress has been made towards de novo protein design using both diffusion models \cite{REF44_watson2023novo} and large language models \cite{REF45_madani2023large}. These approaches have shown promising results in generating novel proteins which conform to imposed spatial constraints (Figure \ref{fig:Horizon2fig3}, `Generative Biology'). Research has seen approximately 15\% success rates in the lab, with examples demonstrating the ability to enhance binding potency by 100 to 1000 times for AI-designed proteins. This remarkable success in research will herald a new era of protein science. There is still a way to go before generative methods design novel therapeutics as there are many more objectives to satisfy beyond spatial structure. However, the field is advancing at a rapid pace, and it is anticipated that generative methods will become increasingly adopted in chemical biology in the near future. Such advances will play a critical role in enabling researchers to treat the untreatable.
\\

\subsection{Advancing imaging for digital pathology}

Imaging plays a role across pharmaceutical R\&D from phenotypic screening with cellular assays in molecular discovery through to digital pathology in the preclinical and clinical stages. Computer vision techniques have been used for some time to make quantitative assessments of samples or automate manual evaluations. The future of generative AI for imaging applications will become increasingly specialist, moving from synthetic data for augmented model training and data privacy, to correcting image quality issues and providing standardisation across samples from different institutions around the globe. This will expand beyond colour
\linebreak
normalisation and allow comparison of samples established using different staining techniques, further increasing the size of data sets for model training and combination of different datasets to generate coherent images containing multiple data layers (Figure \ref{fig:Horizon2fig3}, `Digital Pathology'). This impact will be most significant in the preclinical phase where there is an opportunity to accelerate and improve decision-making at scale. As discussed in the previous section, synthetic data applications need to be approached with caution, but the potential future value is significant.
\\

It has already been shown that generative AI can create high-quality digital pathology images in areas such as oncology \cite{REF46_quiros2019pathologygan}, and in the future the technology will enable synthetic data to be created across resolution scales including whole-slide images at lower resolution, of which there are far fewer samples available \cite{REF47_morrison2021generative}. It must be cautioned that generative AI cannot be allowed to create information where there is none, and careful validation of these novel approaches will be essential. With this in place, these advances will greatly advance preclinical decision-making improving, impacting \textit{speed} and \textit{quality}.
\\

\subsection{More efficient and patient friendly clinical trials}

Operationally, clinical trials are necessarily bureaucratic, with the integrity, safety and
\linebreak
effectiveness all hanging on a clearly defined protocol, a plethora of standard operating
\allowbreak
procedures, which define the process for everyone from patients to physicians and trial 
\linebreak
statisticians to external Contract Research Organizations (CROs). Generative AI offers the potential to support the drafting of new trial protocols subject to the constraints of the trial (Figure \ref{fig:Horizon2fig3}, `Patient experience on clinical trials'). It further could support communication between the different parties engaged in the operation of the clinical trial and be the first choice to write the first draft of a range of documents.
\\

While this may appear to be a high-risk application for a technology which regularly makes errors, the risk is all but eliminated if generative AI is considered as writing a first draft, which will then be reviewed and developed further by a human team. These human-in-the-loop activities lend themselves to generative AI because significant productivity gains are possible without significant risks – even in high stakes applications. Indeed, the partnership between human and machine may enable better outcomes while preserving the high-quality control 
demanded in such applications. In addition, engagement of clinical trial subjects is essential to capturing the high-quality data needed to demonstrate efficacy and safety of novel therapeutics. Beyond requiring high levels of patient adherence for a successful clinical trial, there are wider ethical considerations which dictate that patient burden should be minimised at every stage of a trial \cite{REF48_li2023ethics}.
\\

Generative AI has a role to play in patient experience in enabling trial subjects to be more informed about the trial procedures and their own healthcare. At the simplest levels, this can include purely language-focused tasks such as translating medical jargon to make it accessible to a non-expert audience \cite{REF49_BusinessWireVitalReleasesDoctorToPatientTranslator}. Indeed, the same concepts which can break down internal barriers between specialist teams within the organisation can be adapted to patient communication. Beyond this, there is a range of areas where additional value can come from generative AI. This includes engaging around reminders, answering questions related to appointments, tests and doses, as well as helping with reporting throughout the trial duration. In particular, real-time reporting through patient diaries could be impacted. Transforming the patient experience will not only improve the quality of data from clinical trials, but forms part of a sustainable vision for the future of healthcare.
\\

\subsection{Digital biomarkers and tackling the future burden of disease}

A key strength of generative AI techniques is the ability to learn complex representations of data for a range of different tasks. The ability to learn representations of patient populations is not limited to data modalities such as structured data, text, and images. Valuable digital biomarkers have been developed which leverage AI with data from various sources. This includes everything from wearable devices and the classification and quantification of physical activity \cite{REF50_kourtis2019digital} to the use of audio data to infer disease states from linguistic and acoustic information \cite{REF51_hajjar2023development}.
\\

\begin{figure}[H]
    \centering
    \includegraphics[scale=0.6]{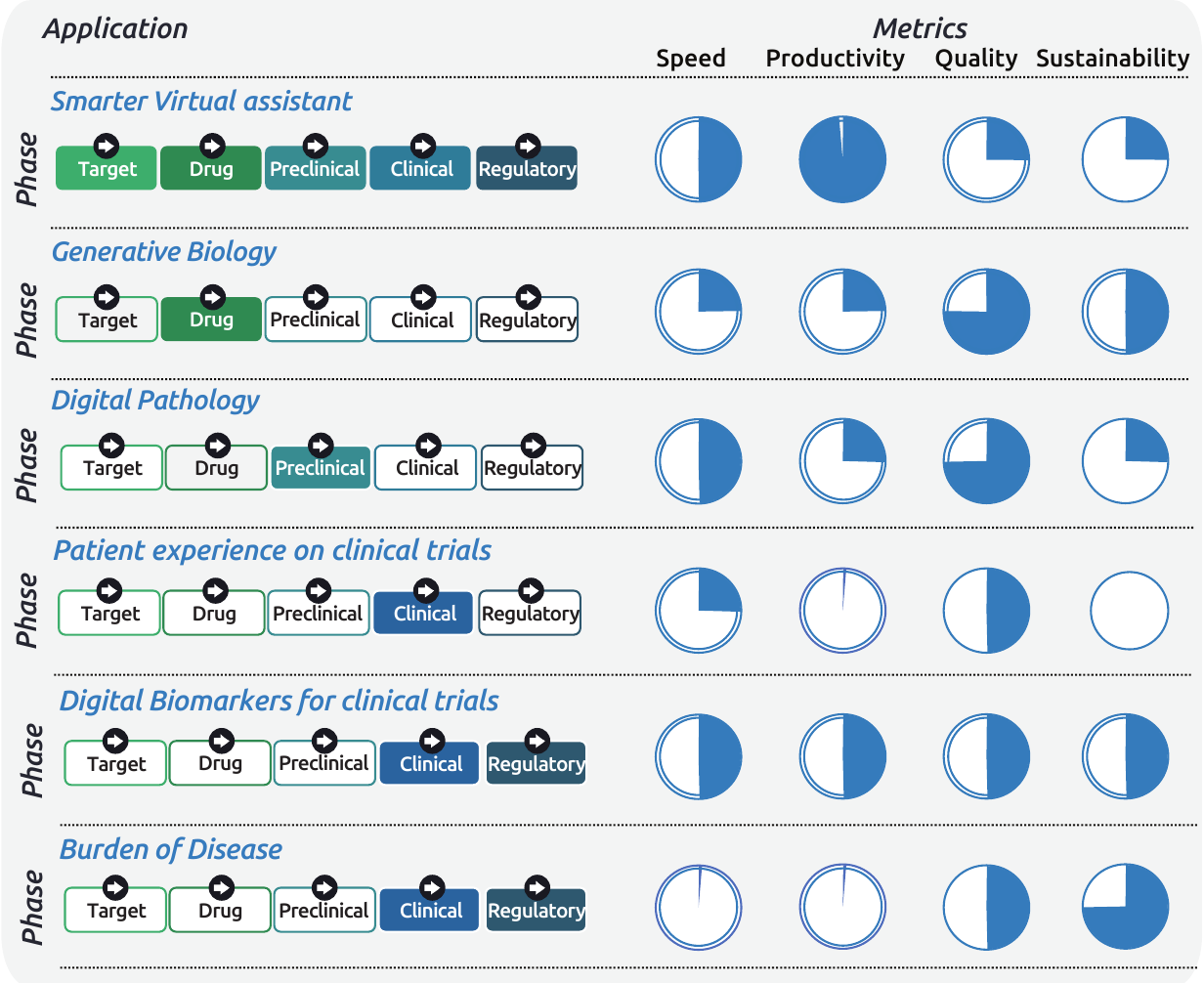}
    \caption{Horizon 2: Suggested use case development timeline, showing key stages and metrics. Speed, productivity, quality, and sustainability are key metrics for suggested use case development.}
    \label{fig:Horizon2fig3}
\end{figure}

Just as traditional biomarkers are frequently multimodal in nature through aggregating multiple clinical and demographic features into risk scores, the future of digital biomarkers is also multimodal \cite{REF52_clay6multimodal}.
Whether it be through data captured from multi-sensor wearable devices or real-world use of digital therapeutics, understanding of patient populations will reach unparalleled depths (Figure \ref{fig:Horizon2fig3}, `Digital biomarkers for clinical trials'). This will be a key enabler of the precision medicine of tomorrow.
\\

As we move further into the future, breakthroughs which lead to the treatments of tomorrow will no longer arise solely from the identification of novel disease targets or potent, selective drug molecules. The role of data in patient care is evolving to enable new modalities of therapeutic intervention which are fundamentally digital in nature. The past few years has seen an acceleration in the number of digital companions and digital therapeutics (DTx) approved for medical use \cite{REF53_wang2023digital}. Such products range from helping Type-2 Diabetes patients to optimise dose base on their personal data, to the use of virtual reality (VR) approaches to back pain relief \cite{REF54_garcia20218}.
\\

The trend for augmenting, personalising and substituting traditional pharmaceuticals with DTx could be accelerated by generative AI techniques (Figure \ref{fig:Horizon2fig3}, `Burden of Disease'). Through the ability learn representations of the space of patient populations, their response to therapeutic interventions, and the likely behaviours of individuals, the needs of more patients could be met. Furthermore, the interactive experience which generative AI will offer patients on clinical trials, discussed above, will eventually find more widespread value post-approval. Such creative approaches to meeting the future health burden will be central to a future vision of sustainable health for all.

\section{Discussion}
\subsection{Horizon 3 - A vision for the future}

Moving beyond the AI technology of tomorrow, an era of increasing bioconvergence 
lies on the horizon, where multiple technologies will intertwine as they mature. The confluence of engineering and computation with biotechnology will lead to a ground-breaking shift in pharmaceutical R\&D which will span the fields of AI, computing, biology, chemistry, and medicine. Interdisciplinary research is at the heart of this transformative process, drawing on multiple breakthroughs from specialized areas to create synergistic solutions. This convergence will be driven by rapid advances in the ability to capture and consume specialised data at scale, spanning areas such as single-cell omics, gene editing and synthetic biology, biosensors, multiplex assays, robotics, materials sciences, and digital twins. AI, generative or otherwise, will be at the centre of this transformation, which will see the timescales for R\&D reduced, the quality of data and decision-making maximised, and focus staff productivity on critical challenges which require human expertise, rather than repetitive process. Looking through the telescope to the next decade and beyond, lies a sustainable industry providing treatments across the range of healthcare needs, meeting environmental goals, and continually investing in strategic innovation to futureproof long-term growth.

\subsection{Closing the loop}

\begin{figure}[H]
    \centering
    \includegraphics[scale=0.6]{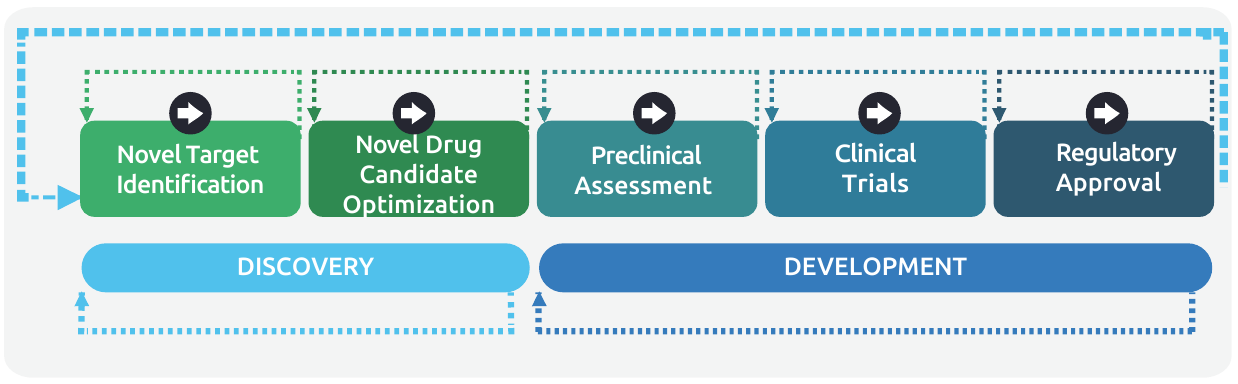}
    \caption{A paradigm shift towards ‘closing the loop’ in drug discovery and development processes, merging computational modelling, AI, and automated high-throughput experiments to rapidly test hypotheses and expedite advancements in patient therapies. This transformative approach not only accelerates scientific experimentation, but also paves the way for a 'lights out' laboratory, automating and linking multiple processes and generating new data at unprecedented speeds.}
    \label{fig:ClosingTheLoopfig4}
\end{figure}

In practical terms, this convergence promises to usher in new therapies, treatment modalities, and innovative ways of enhancing the patient experience. It is poised to transform research and development in each phase, with automation and rapid decision-making at the heart of scientific experimentation and discovery. This paradigm shift revolves around the concept of "closing the loop" in discovery and development processes, where hypotheses are not just formulated but also rigorously tested in a continuous, iterative cycle (Figure \ref{fig:ClosingTheLoopfig4}). Rapid exploration will be possible through the prioritization of in silico experiments.
\allowbreak
Harnessing the power of computational modelling, simulations and AI, a multitude of potential drug candidates and their interactions with human biology will be investigated at a larger scale and faster pace. “Closing the loop” is not limited to in silico experiments alone. Automation of optimally selected high-throughput experiments will complement these efforts, allowing researchers to generate specific and new data at unprecedented speed. The idea of a “lights out” laboratory may be initially focused on specific tasks \cite{REF55_reed2021lab}, but with time expand to link multiple processes together.
\\

Beyond speed and scale will be the sheer richness of data that can be captured. This ranges from advances in imaging technology to capture greater content to being able to probe the complexity of in vivo processes. The ‘language of life’ has been a way of referring to the complexities encoded in DNA across the decades \cite{REF56_berlinski1986language}. 
Genomics has played a huge role in pharmaceuticals from the identification and validation of drug targets to the identification of patients who will benefit from a novel treatment. Understanding the complex processes associated with diseases and potential interventions requires data across the genome, the transcriptome, the proteome, the metabolome and beyond. Indeed, the language of life takes the form of multimodal data – data in different forms combined – and is expressed in terms of multi-omics.
\\

Building on the unprecedented ability to probe biology are techniques which allow the creation of artificial life, such as the development of artificial human tissues, organoid models and organ-on-a-chip technologies which capture physiological processes and functionalities. Not only does this offer the potential for removing animal models from preclinical studies, but may also offer the chance to capture mechanisms not faithfully represented in such studies \cite{REF57_vunjak2021organs}. This potential to bypass traditional animal testing \cite{REF58_ScienceNewPathToNewDrugs} not has the potential to reduce the time to enter clinical trials, but removes ethical concerns associated with such experiments. This further speaks to future sustainability of the industry. The ability to engineer and model biological systems, is already facilitating a range of opportunities spanning from new modalities such as cell therapy \cite{REF59_cappell2023long} to the biomanufacturing of antibodies and other drug molecules \cite{REF60_jozala2016biopharmaceuticals}. Synthetic biology will play an increasingly significant role from early discovery to the clinic, from engineering a new therapeutic agent \cite{REF61_mallick2021bionanomaterials} to the development of biosensors which can detect disease and allow treatments to be better targeted \cite{REF62_wang2023biosensor}.
\\

All these opportunities to gain rapid, rich data will enable AI-driven iteration loops across each stage of the R\&D pipeline. The continual and automated generation and testing of hypotheses will extend from discovery to development. Optimising the collection of data applies just as much to the clinical stages of development as the design of drug molecules. AI will be central to automating the strategy for designing and operating a clinical trial, detecting signals early, optimising engagement and facilitating decentralisation, with a laser focus on the outcome. It will also allow sustainable manufacturing to be planned, and earlier in the discovery process. Indeed, it is possible to see larger closed loop optimisations forming around multiple phases of the R\&D pipeline. In a world where acquisition of data is no longer a bottleneck, significant investment in the connection and communication between processes will be critical. The future of AI is multi-agent, with expert AI systems able to collaborate to optimise the pipeline as a whole. The portfolio of projects itself will be monitored by AI, with risks continually balanced. By closing the loop around the R\&D pipeline, AI will be able to help R\&D leaders identify how to optimise the end-to-end lifecycle, with new innovations achieving impact sooner through the accelerated timelines. This future vision puts AI at the centre of each step of the journey from the lab to the clinic.

\section{Conclusion: Develop a strategy for today}

Given the long-term transformative impact of generative AI technology, R\&D strategy should look to generative AI to support near-term and long-term goals. Getting value from generative AI for R\&D involves balancing a portfolio of investments as one part of an organisations wider R\&D strategy. Such a portfolio should balance getting value today from opportunities in Horizon 1, with investments to shape the disruptive Horizon 2 opportunities of tomorrow. Placing informed bets will be crucial to ensure long-term
\allowbreak
transformative value is unlocked – something which mirrors drug discovery R\&D, where a portfolio approach and fail-fast culture is adopted.
\\

In the near term, this means looking to organising data and identifying early wins from productivity gains across functions. Focussed applications such as small-molecule de novo generation will likely already be delivering value and, looking to the second generation of applications, organisations should look to learn from successes and find similar opportunities across the pipeline. The second wave of targeted opportunities will impact the clinical arena, from operations to patient experience. R\&D organisations should explore the ROI expected from each of these and prepare for the disruption they will bring. Long-term transformation will be catalysed by advances across technologies, and the future synergies should be considered in all technology investments. The opportunities to ‘close the loop’ will likely be seen first in stages of discovery. Given the rapid maturation of generative AI, an up-to-date technology radar should be maintained, which should include these Horizon-3 applications. Some applications will progress more incrementally while delivering increasing value at every stage. An example of this is the virtual assistant, whose abilities will grow as generative AI applications improve and end users drive the focus towards the capabilities which have the biggest impact on their productivity.
\\

A route to piloting generative AI technology in a safe and effective way is through
\linebreak
partnerships with experienced vendors, specialized in scaling up state-of-the-art technologies. However, as generative AI will increasingly become a core capability within R\&D data science teams, organisations should put in place a strategy for developing the future workforce. The
\linebreak
cross-function impact of generative AI means a model for exploration of Proofs of Concepts (PoCs), development of use cases and adoption should be cross-organisational in nature. As generative AI continues to revolutionize pharmaceutical R\&D, from enhancing early drug discovery to reshaping clinical development and beyond, it is imperative for industry leaders to craft forward-thinking strategies that harness this technology to boost productivity, improve research quality, and pave the way for a sustainable and transformative future in treating and curing diseases.

\section*{Acknowledgments}

The authors wish to thank Flavio Morelli and James Hinchliffe for their invaluable review and insightful comments on the manuscript, which significantly contributed to the refinement of this work. We also thank to Aled Diplock for his support in the formatting and restructuring of the manuscript, ensuring its coherence and readability.

\printbibliography

\end{document}